\documentclass[3p,times]{elsarticle}
\usepackage{amsmath}
\usepackage{latexsym}
\usepackage{mathrsfs}
\usepackage{graphicx}
\usepackage{dcolumn}
\usepackage{graphics}
\usepackage{color}
\usepackage{bm}
\usepackage{units,nicefrac}

\begin{document}
	
\begin{frontmatter}

\title{Structure and Morphology of Crystalline Nuclei \\ arising in a Crystallizing Liquid Metallic Film}

\author[kfu]{Bulat N. Galimzyanov}
\ead{bulatgnmail@gmail.com}

\author[kfu]{Dinar T. Yarullin}
\ead{YarullinDT@gmail.com}

\author[kfu,ufrc]{Anatolii V. Mokshin}
\ead{anatolii.mokshin@kpfu.ru}

\address[kfu]{Kazan Federal University, 420008 Kazan, Russia}
\address[ufrc]{Udmurt Federal Research Center of the Ural Branch of the Russian Academy of Sciences, 426067 Izhevsk, Russia}

\date{\today}

\begin{abstract}
Control of the crystallization process at the microscopic level makes it possible to generate the nanocrystalline samples with the desired structural and morphological properties, that is of great importance for modern industry. In the present work, we study the influence of supercooling on the structure and morphology of the crystalline nuclei arising and growing within a liquid metallic film. The cluster analysis allows us to compute the diffraction patterns and to evaluate the morphological characteristics (the linear sizes of the homogeneous part and the thickness of the surface layer) of the crystalline nuclei emergent in the system at different levels of supercooling. We find that the liquid metallic film at the temperatures corresponded to low supercooling levels crystallizes into a monocrystal, whereas a polycrystalline structure forms at deep supercooling levels. We find that the temperature dependence of critical size of the crystalline nuclei contains two distinguishable regimes with the crossover temperature $T/T_{g}\approx1.15$ ($T_{g}$ is the glass transition temperature), which appears due to the specific geometry of the system.
\end{abstract}


\end{frontmatter}

\section{Introduction}
The metallic glasses are materials with unique functional and engineering properties that allow them to find a wide application in the production of high-tech electronics, microprocessors, storage devices and biomedical equipments~\cite{Greer_2009,Kelton_Greer_2010,Berthier_2011,Egami_2015}. As known, metallic glasses are mainly prepared by fast cooling of equilibrated melts to temperatures much below the melting temperature $T_{m}$ avoiding or significantly slowing down crystallization~\cite{Zhong_Mao_2014,Royall_Williams_2015}. The specific optical, magnetic and mechanical properties of metallic glasses depend mainly on their microscopic structure and depend also on the presence of crystalline inclusions~\cite{Kelton_Greer_2010,Cheng_Ma_2011}. So, pure glassy materials with a ``perfect'' disordered structure exhibit improved strength and hardness as compared with crystalline analogues~\cite{Cheng_Ma_2011,Ediger_Harrowell_2012}.

Since the 80s of the last century, an amorphous state of the systems serves as a starting point to generate the nanocrystalline samples with required structural and morphological properties~\cite{Gleiter_2000,Liu_Chen_2008,Sosso_Michaelides_2016,Shibuta_Sakane_2016}. Here an reheating of glassy system and an applied external pressure are used as the basic ways to control the crystal nucleation and growth processes as well as the structure and morphology of crystalline nuclei~\cite{Myerson_2002}. Despite an achieved success in study of phase transformations in glassy materials, there is no clear understanding the influence of supercooling on the structural features of emerging crystalline nuclei as well as on their morphological characteristics such as nucleus size and its shape, structural properties of solid-liquid interface~\cite{Ediger_Harrowell_2012,Miracle_2004,Mishin_Asta_2010}. Direct detection of nanocrystallites inside a crystallized bulk glassy system and evaluation of their morphology are complicated tasks even for modern experimental tools related with X-ray diffraction and transmission electron microscopy~\cite{Carter_Williams_2009}. In this regard, an accurate modeling of nucleation and growth processes allows one to understand better the influence of supercooling on the structure and morphology of crystalline nuclei~\cite{Frenkel_Smit_2001,Nada_2018}. The present study is mainly aimed to consider this issue. Namely, we study the crystallization of liquid metallic film at different levels of supercooling by means of atomistic dynamics simulations. The main attention is paid to study influence of supercooling degree on the structure and morphology of growing crystalline nuclei.

The present work is organized as follows. In Section 2, the simulation details and applied computational methods are presented. Section 3 consists of three subsections. Subsection 3.1 focuses on structure of crystalline nuclei at different levels of supercooling. The temperature dependencies of the main kinetic and thermodynamic characteristics -- the nucleation waiting time and the average size -- of the critically-sized nucleus are discussed in subsection 3.2. The morphology of growing crystalline domains is considered in subsection 3.3. Conclusions are given in Section 4.

\section{Simulation Details and Computational Methods}

\subsection{Details of Considered Dzugutov (Dz) System}

Interatomic interactions in metallic systems are reproduced by different types of model potentials (pair spherical potentials, pseudopotentials, etc.). At present, it is supposed that the most realistic results for metals can be obtained using EAM-potentials. However, it turns out that in some cases the simplified spherical pair potentials are able to provide more realistic results for structural and dynamical properties of metals than EAM-potentials~\cite{Gonzalez_2001,Khustutdinoff_2018,Mokshin_Galimzyanov_2018}. The Dzugutov potential applied to simulate the metallic system in the given study can be considered as one of the simple models of interparticle interaction in liquid metals. This potential is short-ranged and oscillated, that mimics the well-known features of ion-ion interaction influenced by the electron screening effects as well as the Friedel oscillations in the metallic systems. Note that the phase diagram of the Dzugutov system contains all the elements typical for metals (for details, see~\cite{Roth_2000}).

\begin{figure}[ht]
	\begin{center}
		\includegraphics[width=9cm]{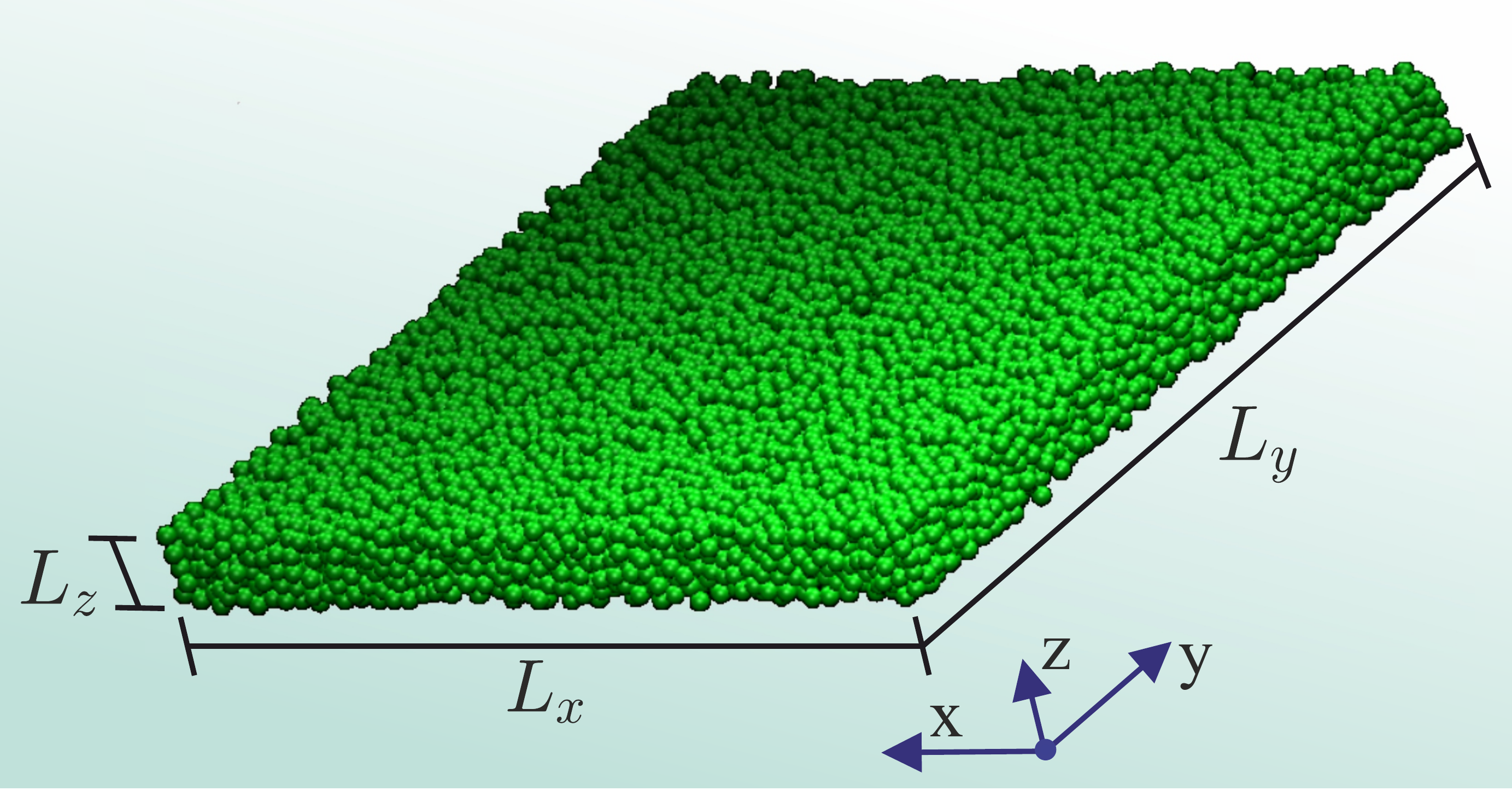}
		\caption{(color online) Simulation cell with $14\,700$ particles. The periodic boundary conditions are applied in the directions $x$ and $y$; and $L_{z}\simeq4.8\,\sigma$ is the width of the film.} \label{fig_1}
	\end{center}
\end{figure}

As known, the necessary condition to study the phase transitions by means of (atomistic) molecular dynamics simulations is that the size of a simulation cell has to exceed the size of emerging new phase nuclei. Fulfillment of this condition allows one to avoid appearance of artifacts related with crystallization induced by periodic boundary conditions. This issue has been considered in detail by Wedekind et al.~\cite{Wedekind_Reguera_2006}; and it was argued that the smallest size of the simulation cell to study nucleation in three-dimensional system is about $5000$ particles. Obviously, if the system has natural size limitation along a direction (for example, as for a film), the size of the simulation cell may be even smaller than Wedekind’s value with $5000$ particles. The simulated system contains $n=14\,700$ identical particles located inside a cell of the volume $V=L_{x}\times L_{y}\times L_{z}$ with $L_{x}=L_{y}\simeq11.5\,L_{z}$ and $L_{z}\simeq4.8\,\sigma$~\cite{Galimzyanov_Yarullin_Mokshin_2018}. The periodic boundary conditions are applied only in the directions $x$ and $y$ (see Figure~\ref{fig_1}). The particles interact through the pair Dzugutov potential~\cite{Roth_2000,Dzugutov_1992}, which promotes the formation of a relatively stable amorphous state. The equations of motion are integrated with the time step $\Delta t=0.005\,\tau$.

The simulations are performed in the isobaric-isothermal ensemble, where the system temperature $T$ and the pressure $P$ are controlled by means of the Nose-Hoover thermostat and barostat, respectively, with optimal coupling constants~\cite{Frenkel_Smit_2001,Evans_Holian_1985}. The fast isobaric cooling from well equilibrated liquid with the temperature $T=3.0\,\epsilon/k_{B}$ is applied. The supercooled samples are generated for the thermodynamic states with the pressure $P=15\,\epsilon/\sigma^{3}$ and at the temperatures within the range from $T=0.5$ to $1.4\,\epsilon/k_{B}$ that are below of the melting temperature $T_{m}\simeq1.72\,\epsilon/k_{B}$~\cite{Roth_2000}. The glass-transition temperature of the system $T_{g}$ was determined through estimation of the Wendt-Abraham parameter~\cite{Wendt_Abraham_1978} and it was found that at the cooling rate $0.04\,\epsilon/(k_{B}\tau)$ and the applied pressure $P=15\,\epsilon/\sigma^{3}$ the glass transition temperature is $T_{g}\simeq0.78\,\epsilon/k_{B}$. Thus, this study covers the states with the supercooling levels from $(T_{m}-T)/T_{m}\simeq0.19$ (at the temperature $T=1.4\,\epsilon/k_{B}$) to $(T_{m}-T)/T_{m}\simeq0.71$ (at the temperature $T=0.5\,\epsilon/k_{B}$).

For convenience, all the quantities in this work are expressed in terms of the parameters of the potential: the effective particle diameter $\sigma$ and the energy parameter $\epsilon$. Then, the time unit is $\tau=\sigma\sqrt{m/\epsilon}$, where $m$ is the particle mass; the wave number unit is $1/\sigma$; the temperature unit is $\epsilon/k_{B}$. Consequently, for a real metallic system, the numerical results of this study can be expressed in absolute units. For example, for the case of iron with the parameters $\sigma\simeq2.52$\,\AA~and $\epsilon\simeq16.2$\,kcal/mol~\cite{Bykata_Borges_2015}, we have the time $\tau\simeq0.2$\,ps as well as the time step $\Delta t\simeq1\,$fs (at $\Delta t=0.005\,\tau$).

\subsection{Cluster Analysis and Structural Analysis of Instantaneous Configurations} We perform the standard cluster analysis to recognize crystalline structures on the basis of the instantaneous configurations of the simulated samples. Namely, the local orientational order parameters are computed for each particle of the system to identify the ``solid-like'' particles~\cite{Steinhardt_1983,Mickel_Mecke_2013}. Then, the ten-Wolde-Frenkel scheme is applied, according to which a ``solid-like'' particle is identified as particle incoming in a crystalline nucleus if the particle contains not less than $7$ ``solid-like'' neighbors~\cite{Wolde_Frenkel_1996}. Details of the cluster analysis algorithm are given in Refs.~\cite{Mokshin_Galimzyanov_2015,Mokshin_Galimzyanov_2017}.

Information about the structure of crystalline domains as well as about a type of crystal lattice symmetry and its spatial orientation are obtained through evaluation of the structure factor projected on the $xy$-plane~\cite{Egami_2015,Chekmarev_2001}:
\begin{equation}\label{eq_structure_factor_1}
S(k_{x},k_{y})=\langle\rho(k_{x},k_{y})\rho^{*}(k_{x},k_{y})\rangle,
\end{equation}
where
\begin{equation}\label{eq_structure_factor_2}
\rho(k_{x},k_{y})=\frac{1}{\sqrt{n}}\sum_{j=1}^{n}e^{-i(k_{x}r_{xj}+k_{y}r_{yj})}.
\end{equation}
Here, $k_{x}$ and $k_{y}$ are the components of the wave vector $\vec{k}=\{k_{x},\,k_{y},\,k_{z}\}$; $\rho(k_{x},k_{y})$ is the Fourier-component of the local density fluctuations~\cite{Balucani_Zoppi_1994,Hansen_McDonald_2006,Kryuchkov_Ryzhov_2018}; $r_{xj}$ and $r_{yj}$ are the radius-vector components of the $j$th particle. The angle brackets $\langle...\rangle$ mean the ensemble averaging. Note that, at evaluation of $S(k_{x},k_{y})$, the $z$-component of the wave vector $k_{z}$ is not taken into account.
\begin{figure*}[ht]
	\begin{center}
		\includegraphics[width=14.0cm]{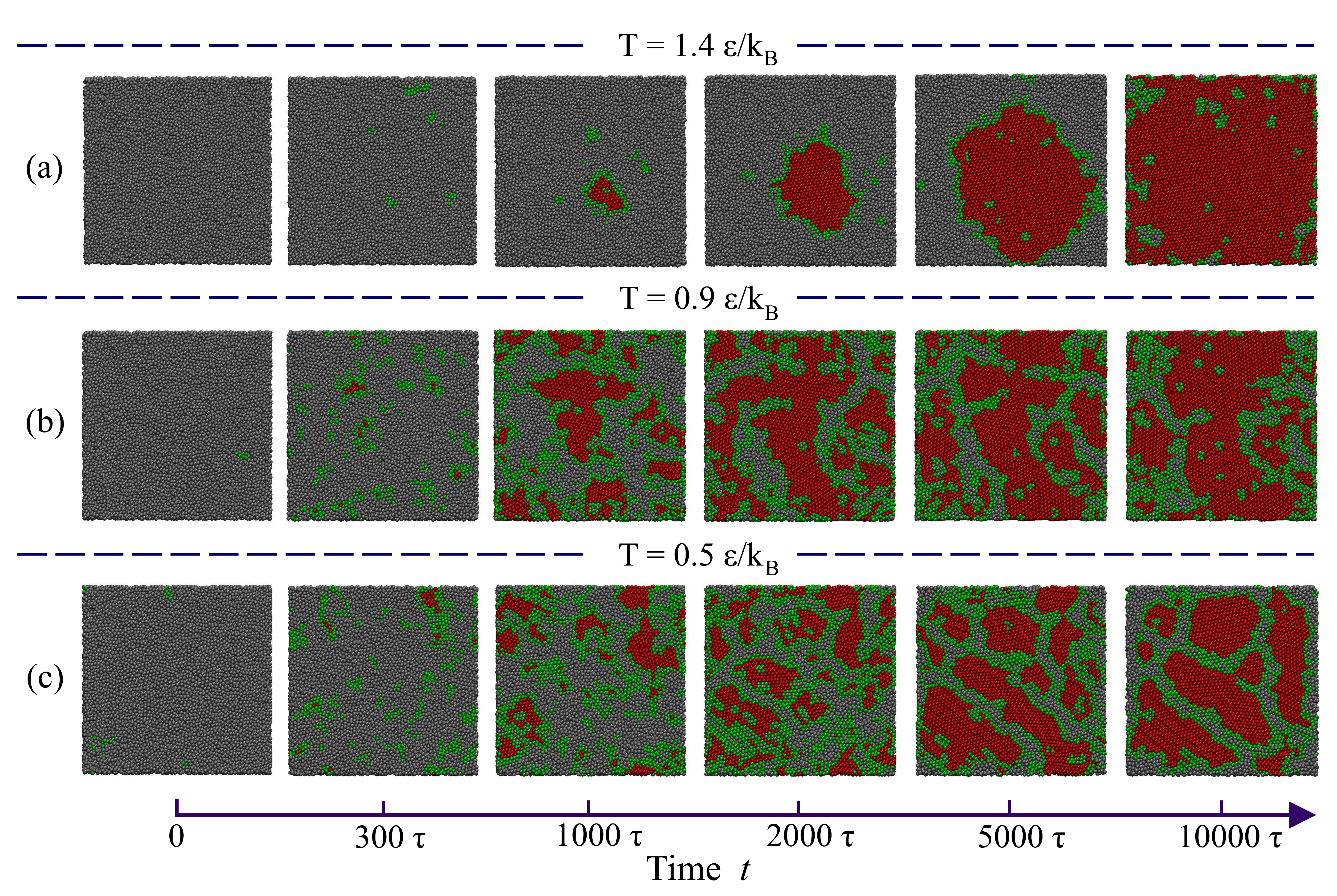}
		\caption{(color online) Snapshots of the system at different time moments from start of crystallization and at different temperatures: (a) $T=1.4\,\epsilon/k_{B}$, (b) $T=0.9\,\epsilon/k_{B}$, and (c) $T=0.5\,\epsilon/k_{B}$. The solid-like particles of crystalline structures are marked by red color, whereas the surface particles are marked by green color. The particles, which form the disordered phase are marked by gray color.}
		\label{fig_2}
	\end{center}
\end{figure*}

\subsection{Estimation of Nucleation and Morphological Characteristics} For each considered ($P$, $T$)-state, the critical size $n_{c}$ and the waiting time $\tau_{c}$ for the first (largest) crystalline nucleus are estimated using the mean-first-passage-time method realized on the basis of the cluster analysis results. Details of this method are given in Refs.~\cite{Mokshin_Galimzyanov_2015,Mokshin_Galimzyanov_2012,Dellago_2016}.

In addition, we focus on study of the morphological characteristics (such as the nucleus radius, the radius of nucleus homogeneous part and the thickness of nucleus surface layer) of the nuclei whose size is equal to and larger than the critical size $n_{c}$. For quantitative evaluation of these characteristics, we determine the number of particles $N$ which generate a nucleus and are inside a sphere of the radius $R$. Note that the geometric center of this sphere coincides with the geometric center of the nucleus. Thus, we determine the correspondence between the number $N$ of particles enclosed within a sphere of the radius $R$
\begin{equation}\label{eq_fractality_1}
N=\left[\frac{R}{R_{0}}\right]^{f},
\end{equation}
while value of the radius $R$ increases from zero to $L_{x}/2$ with the step $R_{0}=0.5\,\sigma$ that is the radius of a particle. Here, $f$ is the fractal parameter, which can take non-integer values within the range from $1$ (for one-dimensional, linear structure) to $3$ (for three-dimensional structure); whereas for the two-dimensional clusters on has $f=2$~\cite{Nigmatullin_Alekhin_2006}. Thus, the morphology of the crystalline nucleus can be evaluated quantitatively from the found correspondence between $\log[N]$ and $\log[R/R_{0}]$.

Further, the average linear size of the crystalline nucleus is estimated as
\begin{equation}\label{eq_average size}
r_{c}=\left(\frac{3N_{cl}}{4\pi\rho}\right)^{1/3},
\end{equation}
where $N_{cl}$ is the number of all particles within the crystalline nucleus, $\rho$ is the numerical density of crystalline phase.

\section{Results and Discussion}

\subsection{Structure of Crystalline Nuclei}

By means of the cluster analysis, the solid-like particles forming the crystalline structures were identified. As an example, some of these structures are presented in Figure~\ref{fig_2}. Namely, Figure~\ref{fig_2}(a) shows snapshots of the system at the temperature $T=1.4\,\epsilon/k_{B}$ corresponding to the thermodynamic state at the lowest considered supercooling level. In Figure~\ref{fig_2}(b), the system configurations at the temperature $T=0.9\,\epsilon/k_{B}$ are presented. This temperature is close to the glass transition temperature $T_{g}\simeq0.78\,\epsilon/k_{B}$ and corresponds to a moderate supercooling level. Figure~\ref{fig_2}(c) presents the configurations at the temperature $T=0.5\,\epsilon/k_{B}$ corresponding to the deepest considered level of supercooling. It can be seen from Figure~\ref{fig_2}(a), when the supercooling is insignificant and the temperature is $T=1.4\,\epsilon/k_{B}$ then one stable nucleus in the simulation cell is formed only. The growing crystalline nucleus has a relatively smoothed interfacial boundary. These features are typical for mononuclear scenario of crystallization that is well agreed with the classical nucleation theory (CNT)~\cite{Dellago_2016,Kashchiev_Nucleation_2000,Huitema_Human_2000,Saika-Voivod_2015}. Moreover, with increase of the supercooling level, decrease of the nuclei size and increase of the nuclei concentration are observed. For example, the ramified crystalline structures formed at the temperature $T=0.9\,\epsilon/k_{B}$ are clear recognized in the snapshots given in  Figure~\ref{fig_2}(b). The coalescence process of these nuclei proceeds through the so-called mechanism of oriented attachment~\cite{Ivanov_Osiko_2014}, and the nuclei coalesce to each other and form a common crystal lattice. Eventually, at the time $t\approx10^{4}\,\tau$, the system generates a monocrystal. This monocrystal contains a small number of defects. In the case of deep supercooling, we observe a more high concentration of ordered domains [see Figure~\ref{fig_2}(c)]. The high concentration of these domains leads to formation of the ramified structures that had been previously erroneously interpreted as signs of the spinodal decomposition~\cite{Sosso_Michaelides_2016,Trudu_Donadio_2006,Bartell_Wu_2007}.

\begin{figure*}[ht]
	\begin{center}
		\includegraphics[width=16cm]{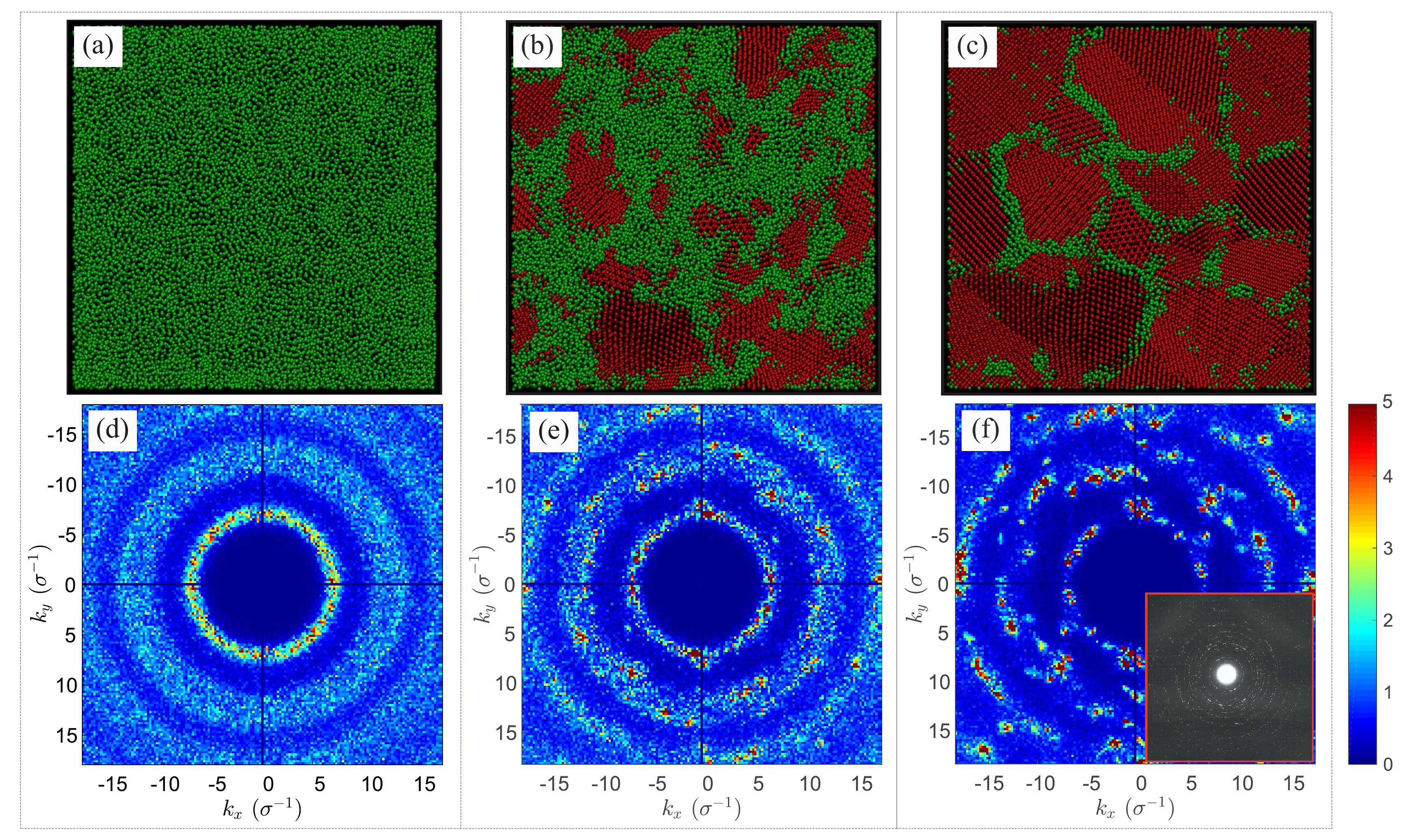}
		\caption{(color online) Top panels: snapshots of the system at the temperature $T=0.5\,\epsilon/k_{B}$ and at different times from start of crystallization: (a) glassy system without the crystallization centers; (b) two-phase state of the system containing crystalline nuclei; (c) polycrystalline system consisting of separate crystalline domains. Bottom panels: $S(k_{x},k_{y})$-projections of the glassy system (d), the two-phase system (e) and the polycrystalline system (f). Inset to panel (f) shows the diffraction pattern obtained by transmission electron microscopy for polycrystalline foil~\cite{Carter_Williams_2009}.}
		\label{fig_3}
	\end{center}
\end{figure*}

In Figure~\ref{fig_3}, snapshots of the instantaneous configurations [Figures~\ref{fig_3}(a), \ref{fig_3}(b) and \ref{fig_3}(c)] and the corresponding $S(k_{x},k_{y})$-projections [Figures~\ref{fig_3}(d), \ref{fig_3}(e) and \ref{fig_3}(f)] are presented for the system at the temperature $T=0.5\,\epsilon/k_{B}$. As seen from Figure~\ref{fig_3}(d), the contour of $S(k_{x},k_{y})$ is characterized by the pronounced rings, that is typical for a ``perfect'' glass. Figure~\ref{fig_3}(b) shows the typical two-phase system at the stage of crystal nucleation. Here, the crystalline nuclei spread uniformly throughout whole volume and approximately $50$\% particles are involved to form the crystalline phase. The $S(k_{x},k_{y})$ depicted on Figure~\ref{fig_3}(e) and corresponded to this system contains randomly distributed spots. Such location of the spots is indication of absence of a preferred spatial orientation of the nuclei crystal lattice. Low intensity of these spots in $S(k_{x},k_{y})$ are due to small size of the nuclei. Figure~\ref{fig_3}(c) shows instantaneous configuration of the polycrystalline system which is formed at the time $t\approx10^{4}\,\tau$. This system represents an ensemble of crystalline domains with different crystal lattice orientations. The corresponded $S(k_{x},k_{y})$ is presented in Figure~\ref{fig_3}(f); where one can observed the pronounced and randomly scattered spots on the background of the barely visible rings. It is remarkable that these features of $S(k_{x},k_{y})$ are identical to the diffraction pattern observed for polycrystalline foil~\cite{Carter_Williams_2009} by transmission electron microscopy (TEM) [see inset of Figure~\ref{fig_3}(f)].

\begin{figure}[ht]
	\begin{center}
		\includegraphics[width=10cm]{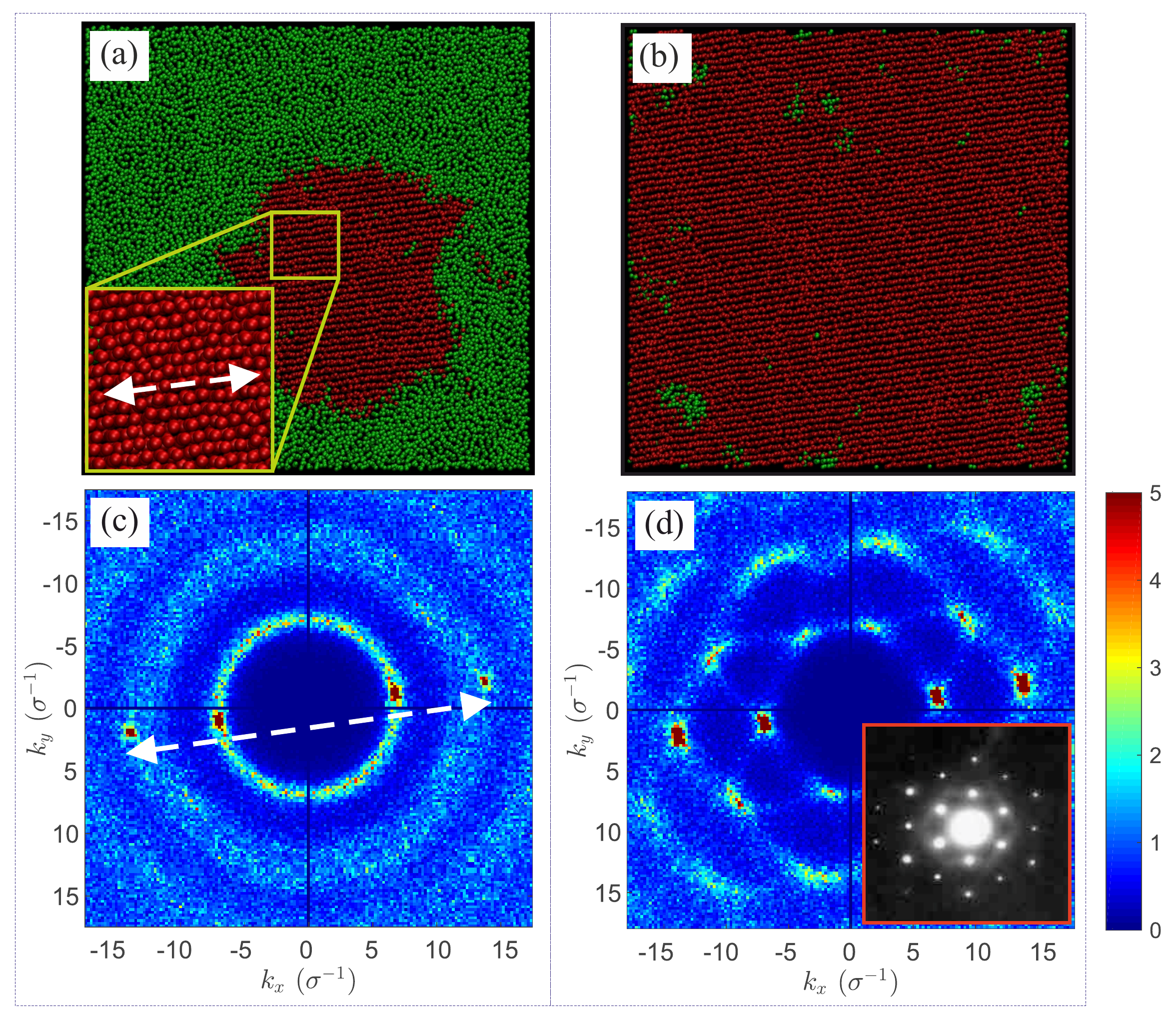}
		\caption{(color online) Top panels: snapshots of the system at the temperature $T=1.4\,\epsilon/k_{B}$: (a) two-phase state of the system containing one crystalline nucleus; (b) monocrystal. Bottom panels: $S(k_{x},k_{y})$-projections of the two-phase system (c) and monocrystal (d). Inset on panel (d) shows the diffraction pattern obtained by transmission electron microscopy for aluminum single crystal~\cite{Carter_Williams_2009}.}
		\label{fig_4}
	\end{center}
\end{figure}

Figure~\ref{fig_4} shows snapshots of the system [Figures~\ref{fig_4}(a) and \ref{fig_4}(b)] and $S(k_{x},k_{y})$-projections [Figures~\ref{fig_4}(c) and \ref{fig_4}(d)] at the temperature $T=1.4\,\epsilon/k_{B}$. At the given temperature, the system crystallizes through formation and growth of a single nucleus, and its shape is characterized by spherical envelope surface [see Figure~\ref{fig_4}(a)]. Here, the structural factor $S(k_{x},k_{y})$ contains pronounced spots located near the rings in accordance with a certain order, and the radii of these rings correspond to the wave numbers $k\simeq2\pi/\sigma$ and $k\simeq4\pi/\sigma$. Such location of the spots is result of a specific crystal lattice of the crystalline domains visible in Figure~\ref{fig_4}(a). Figure~\ref{fig_4}(b) shows snapshot of the system at the time moment, when the growth of the nucleus depicted in Figure~\ref{fig_4}(a) is finished. As seen, the system represents a monocrystal. In this case, the rings in $S(k_{x},k_{y})$ disappear completely. The six-fold symmetry of the diffraction pattern $S(k_{x},k_{y})$ corresponds to crystalline structure with face-centered cubic lattice (fcc)~\cite{Galimzyanov_Yarullin_Mokshin_2018}. This is clearly confirmed by comparision with the diffraction pattern obtained by TEM for a single crystal of aluminum with fcc-structure; this diffraction pattern is also given for comparison in inset of Figure~\ref{fig_4}(d)~\cite{Carter_Williams_2009}. As seen, the simulated $S(k_{x},k_{y})$ and TEM-image obtained for monocrystal are very similar.

\subsection{Nucleation waiting time and critical size of the crystalline nucleus}

\begin{figure}[ht]
	\begin{center}
		\includegraphics[width=8.5cm]{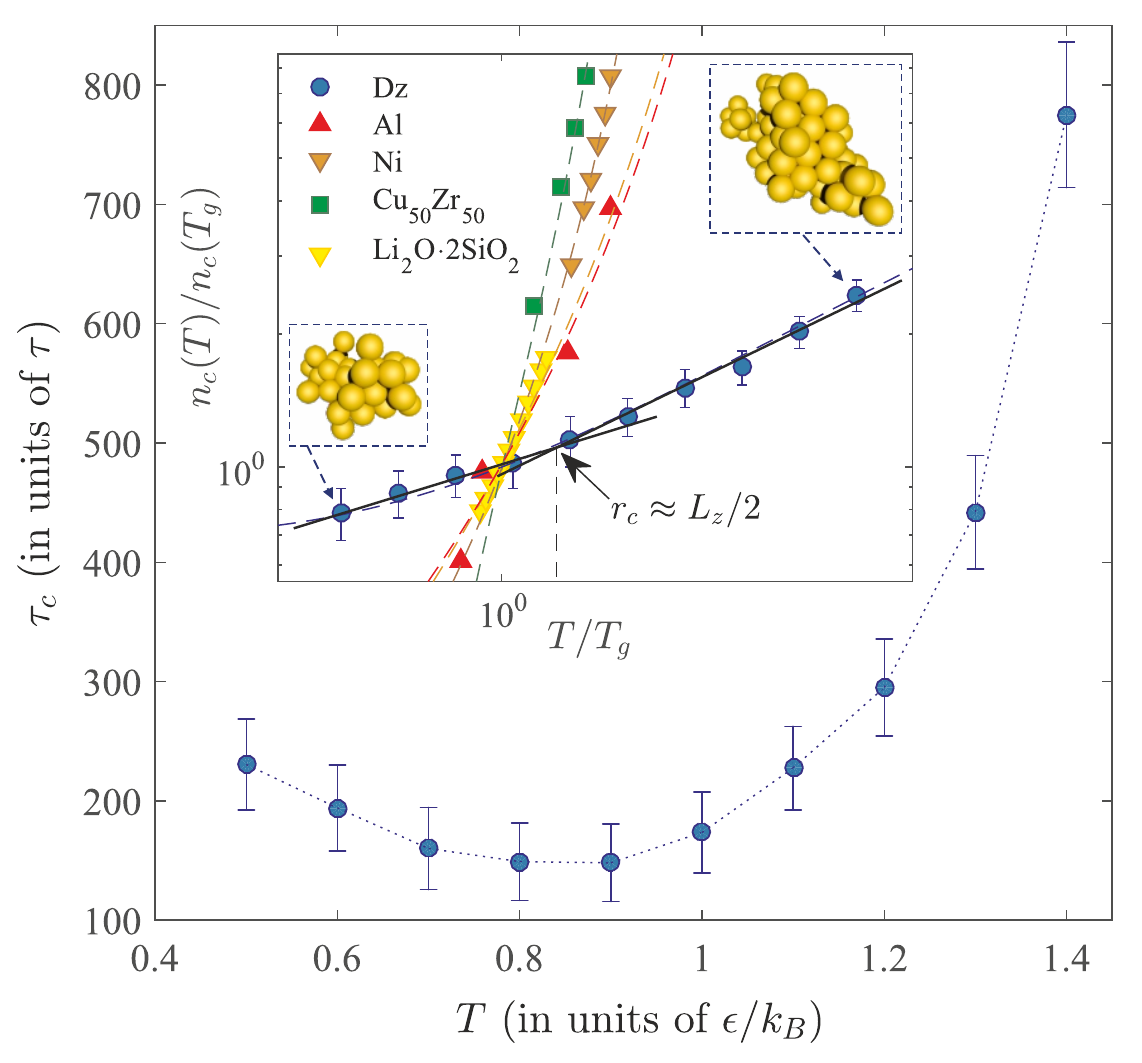}
		\caption{(color online) Waiting time $\tau_{c}$ of the critically-sized nucleus as function of the temperature $T$. Inset: ($T/T_{g}$)-dependence of the reduced critical size $n_{c}(T)/n_{c}(T_{g})$ in double logarithmic scale [where $n_{c}(T_{g})$ is the critical size at the glass transition temperature $T_{g}$] is obtained at crystallization of Dzugutov system (Dz)~\cite{Mokshin_Galimzyanov_2017}. Our results are compared with simulation and experimental data obtained at crystallization of aluminum (Al)~\cite{Mahata_Zaeem_2018}, nickel (Ni)~\cite{Sun_Zhang_2018}, binary melt Cu$_{50}$Zr$_{50}$~\cite{Sato_Ogata_2017} and lithium disilicate Li$_{2}$O$\cdot2$SiO$_{2}$~\cite{Fokin_Potapov_2005}. Dashed curves are results of fit with Eq.~(\ref{eq_nc}). The vertical dashed line indicates the temperature $T/T_{g}\approx1.15$ ($T\approx0.9\,\epsilon/k_{B}$, in absolute units) of the crossover, at which the nucleus average linear size of the critically-sized nucleus becomes comparable to the width $L_{z}$ of the film.}
		\label{fig_5}
	\end{center}
\end{figure}

We have evaluated the nucleation waiting time $\tau_{c}$ and the critical size $n_{c}$ of the nucleus, which are usually considered as the main characteristics of the nucleation. Figure~\ref{fig_5} shows the temperature dependencies of waiting time $\tau_{c}$ and critical size $n_{c}$ (see Table~\ref{Tab_1}). So, $\tau_{c}(T)$ contains the minimum at the temperature $T\simeq0.9\,\epsilon/k_{B}$ corresponded to the largest value of the nucleation rate. Such the nonmonotonic character of the temperature-dependent waiting time is also observed in results for homogeneous crystal nucleation in bulk supercooled systems~\cite{Mokshin_Galimzyanov_2015,Fokin_Zanotto_2006}, and it is resulted from competition of thermodynamic and kinetic nucleation drivers. At the lowest considered supercooling level with the temperature $T=1.4\,\epsilon/k_{B}$, the waiting time takes the largest value, $\tau_{c}=(775\pm61)\,\tau$. Then, this quantity is $\tau_{c}=(230\pm40)\,\tau$ at the temperature $T=0.5\,\epsilon/k_{B}$, which corresponds to a state with a deep supercoolong.
\begin{table}[ht!]
	\caption{Nucleation characteristics of the simulated system: the temperature $T$, the numerical density of crystalline phase $\rho$, the critical size $n_{c}$, the average radius of critically-sized nucleus $r_{c}$ and the nucleation waiting time $\tau_{c}$ are evaluated using simulation data.}\vskip 0.5 cm
	\begin{center}
		\begin{tabular}{ccccc}
			\hline
			$T$ ($\epsilon/k_{B}$) & $\rho$ ($\sigma^{-3}$) & $n_c$ & $r_{c}$ ($\sigma$) & $\tau_c$ ($\tau$)  \\
			\hline
			$0.5$  & $0.946$ & $39\pm5$ & $2.14\pm0.09$ & $231\pm38$ \\
			$0.6$  & $0.925$ & $41\pm5$ & $2.19\pm0.09$ & $194\pm36$ \\
			$0.7$  & $0.909$ & $45\pm5$ & $2.28\pm0.08$ & $160\pm35$ \\
			$0.8$  & $0.898$ & $48\pm6$ & $2.34\pm0.09$ & $149\pm33$ \\
			$0.9$  & $0.889$ & $54\pm7$ & $2.44\pm0.09$ & $148\pm33$ \\
			$1.0$  & $0.885$ & $61\pm6$ & $2.53\pm0.08$ & $174\pm34$ \\
			$1.1$  & $0.883$ & $71\pm7$ & $2.68\pm0.08$ & $228\pm35$ \\
			$1.2$  & $0.882$ & $79\pm7$ & $2.78\pm0.08$ & $295\pm41$ \\
			$1.3$  & $0.881$ & $90\pm8$ & $2.89\pm0.08$ & $442\pm48$ \\
			$1.4$  & $0.880$ & $98\pm10$ & $2.99\pm0.09$ & $775\pm61$ \\
			\hline
		\end{tabular}
		\label{Tab_1}
	\end{center}
\end{table}

The Gibbs free energy $\Delta G$ is the key thermodynamic parameter, and it is a measure of the reversible work necessary for the formation of a nucleus of the size $N_{cl}$~\cite{Kelton_Greer_2010,Kashchiev_Nucleation_2000,Schmelzer_Abyzov_2019,Tipeev_2018}. Thus, the quantity $\Delta G$ is a function of the nucleus size $N_{cl}$. In the case of the homogeneous nucleation, the free energy $\Delta G(N_{cl})$ is defined by negative bulk ($\sim N_{cl}$) and positive surface ($\sim N_{cl}^{2/3}$) contributions:
\begin{equation}\label{CNT_energy}
\Delta G(N_{cl})=\Delta G_{n_c}\left[3\left(\frac{N_{cl}}{n_{c}}\right)^{2/3}-2\left(\frac{N_{cl}}{n_{c}}\right)\right].
\end{equation}
Here, $\Delta G_{n_c}$ is the free energy required to generate the critically-sized nucleus $n_c$. Then, the critical size $n_c$ can be estimated through the following generalized relation
\begin{equation}\label{eq_nc}
n_{c}=C\left(\frac{E_{\gamma}}{|\Delta\mu|}\right)^{3},
\end{equation}
where $E_{\gamma}$ is the interfacial free energy per one particle, $|\Delta\mu|$ is the difference of the chemical potential between the liquid and crystalline phases, and $C$ is the generalized shape-factor of the critically-sized nucleus. Thus, the critical size $n_{c}$ is defined by thermodynamic state of the system. The chemical potential difference $|\Delta\mu|$ in Eq.~(\ref{eq_nc}) reflects a role of thermodynamic aspect in the nucleation process. In addition, the critical size $n_{c}$ is also determined by $E_{\gamma}$ and $C$, whose values strongly depend on the shape of the nucleus. Moreover, the specific geometry of the system has direct impact on the nucleation process, when the linear size of the nucleus is comparable to the size of the system.

There is no rigorous universal temperature dependence $n_c(T)$ and the parameters in relation~(\ref{eq_nc}) -- the interfacial free energy $E_{\gamma}$ and the chemical potential difference $|\Delta\mu|$ -- as functions of temperature are reproduced by different models for the systems of different type. Nevertheless, it is expected~\cite{Sun_Zhang_2018} that relation
\begin{equation}\label{eq_nc_approx}
n_{c}(T)=C_{1}+C_{2}\left(\frac{T_{m}}{T_{m}-T}\right)^{3},
\end{equation}
can be used as the simplest approximation for the case of homogeneous nucleation; here, $C_{1}$ and $C_{2}$ are constants which take certain values for a particular system. The change of the critical size may differ by orders of magnitude for different systems even for the same supercooling range. This effect is determined directly by thermodynamic aspects of nucleation as follows from relation~(\ref{eq_nc}).
\begin{figure}[ht]
	\begin{center}
		\includegraphics[width=8.5cm]{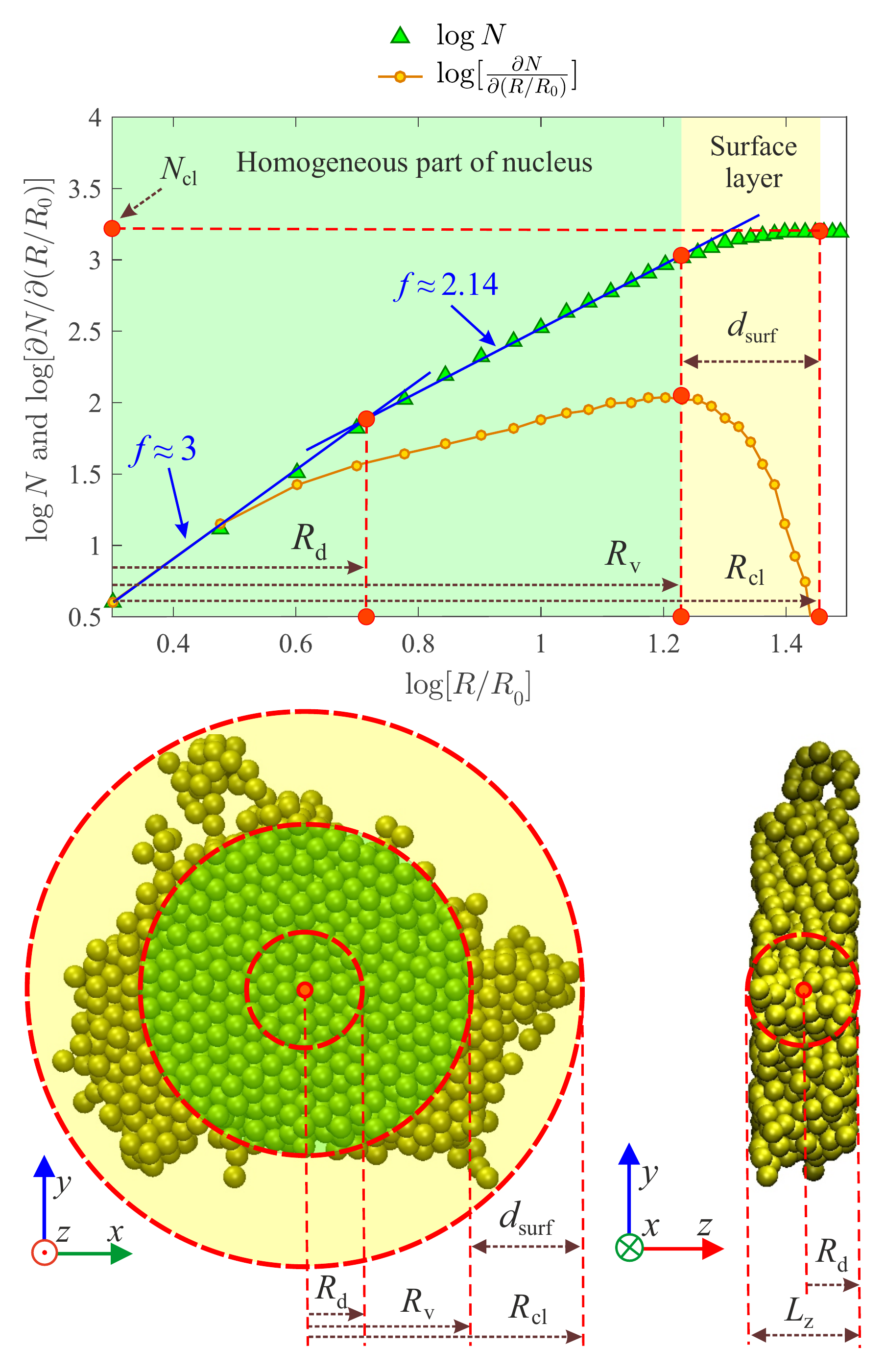}
		\caption{(color online) Top panel: solid curve is $N(R/R_{0})$ and dashed curve is the derivative $\partial N/\partial(R/R_{0})$. Bottom panel: crystalline nucleus of the size $N_{cl}=1568$ particles and of the radius $R_{cl}\simeq14\,\sigma$. For this nucleus, the core spherical region of the radius $R_{d}$, the homogeneous part of the linear size $2R_{v}$ and thickness of the surface layer $d_{surf}$ are shown.}
		\label{fig_6}
	\end{center}
\end{figure}

In inset of Figure~\ref{fig_5}, we present $n_{c}(T)$ obtained for the Dz-system, and compare with available results for the bulk homogeneous crystal nucleation in some systems. Namely, we put in this figure the simulation results for aluminum melt (Al)~\cite{Mahata_Zaeem_2018}, nickel melt (Ni)~\cite{Sun_Zhang_2018} and binary Cu$_{50}$Zr$_{50}$ melt~\cite{Sato_Ogata_2017} and experimental data for lithium disilicate Li$_{2}$O$\cdot2$SiO$_{2}$~\cite{Fokin_Potapov_2005}. As seen, the magnitude of the change of the critical size over the considered temperature range differs for different systems. Nevertheless, all the presented results are well reproduced by common relation~(\ref{eq_nc_approx}). Remarkably, for the Dz-system we find that $n_{c}(T)$ contains two distinguishable regimes with the crossover temperature $T/T_{g}\approx1.15$ ($T\approx0.9\,\epsilon/k_{B}$, in absolute units). These regimes are clearly recognized by log-log scaling applied in the plot of Figure~\ref{fig_5} (inset). For the low temperature range $T/T_{g}<1.15$, the average linear size of the critically-sized nucleus is less than the width $L_{z}$ of the film. At the crossover temperature $T/T_{g}\approx1.15$, the nucleus average linear size becomes comparable to the width $L_{z}$, whereas it is larger than $L_{z}$ at the temperatures $T/T_{g}>1.15$. Thus, this crossover appears due to the specific geometry of the system.

\subsection{Morphology of Crystalline Domains}

Let us now consider the morphological characteristics of crystalline nuclei (domains) whose size is equal to and larger than the critical size $n_{c}$. Such the morphological characteristics of the crystalline nucleus as the nucleus radius $R_{cl}$, the radius $R_{v}$ of homogeneous part of the nucleus and the thickness of surface layer $d_{surf}$ were evaluated according to the scheme presented in Figure~\ref{fig_6}.

This schematic Figure shows the evaluated correspondence between the number $N$ of particles enclosed within a sphere of the increasing radius $R$, that is obtained for the case of the nucleus of the size $N_{cl}=1568$ particles and whose average linear radius is $R_{cl}\simeq14\,\sigma$. As seen, the obtained correspondence $\log[N]$ vs $\log[R/R_{0}]$ consists of two linear parts, the slopes of which are characterized by the fractal parameter $f$ with values $f\approx3$ and $f\approx2.14$. Crossover point between these linear parts is located at $R=R_{d}$, where $R_{d}$ is the radius of the nucleus core range. The fractal parameter $f$ with value $3$ is typical for the core region of the nucleus, whose radius $R_{d}$ equals to half of the simulation cell length, $R_{d}\simeq L_{z}/2$ [see bottom panel on Figure~\ref{fig_6}]. The fractal parameter $f$ with value $2.14$ corresponds to homogeneous part of the nucleus with radius $R_{v}$. The correspondence $\log[N]$ vs $\log[R/R_{0}]$ goes to saturation when $R$ reaches the upper boundary of the nucleus. Thus, the start of the plateau determines the nucleus radius $R_{cl}$. We emphasize that the quantity $d_{surf}$ is not thickness of real solid-liquid boundary layer, but is the thickness of surface layer defined as the difference $d_{surf}=R_{cl}-R_{v}$, where $R_{cl}$ is the radius of envelope circle of the nucleus, and $R_{v}$ is a measure of homogeneous part of the nucleus [see schematic Figure~\ref{fig_6}]. It is clear that the quantity $d_{surf}$ depends directly on the surface roughness. Correct evaluation of the quantities $R_{cl}$ and $R_{v}$ can be done on the basis of analysis of the derivative $\partial N/\partial(R/R_{0})$. As seen from Figure~\ref{fig_6}, the $R/R_{0}$-dependence of the curve $\partial N/\partial(R/R_{0})$ has a pronounced maximum at $R=R_{v}$ and the curve falls to zero at $R\geq R_{cl}$.
\begin{figure*}[ht]
	\begin{center}
		\includegraphics[width=14.5cm]{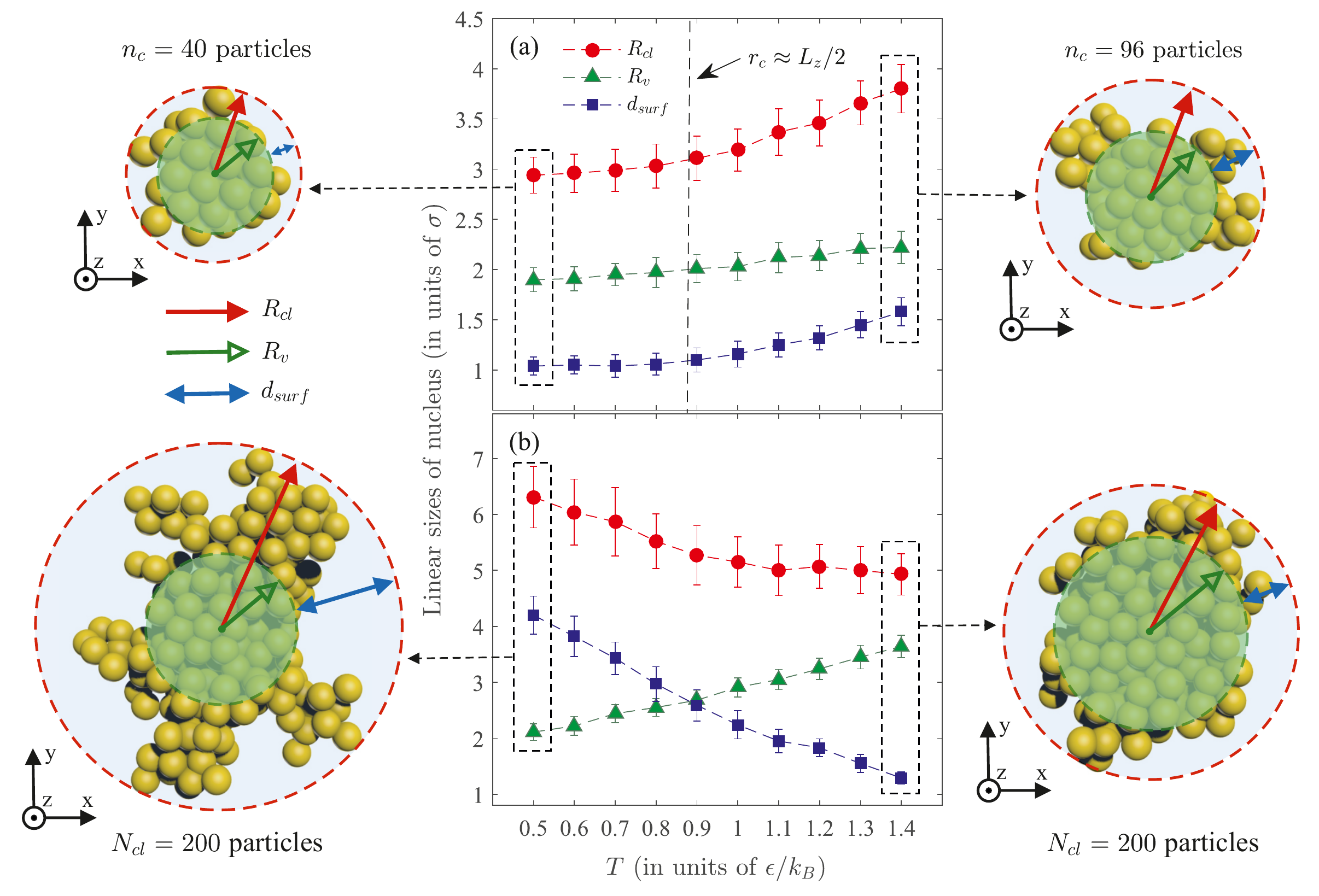}
		\caption{(color online) Temperature dependencies of the nucleus radius $R_{cl}$, the radius of homogeneous part $R_{v}$ and the thickness of surface layer $d_{surf}$: for the critically-sized nucleus (a) and for the nucleus containing $N_{cl}=200$ particles (b). Side insets show the critically-sized clusters and $N$-sized crystalline domains ($N_{cl}=200$ particles) at the temperatures $T=0.5\,\epsilon/k_{B}$ and $T=1.4\,\epsilon/k_{B}$.}
		\label{fig_7}
	\end{center}
\end{figure*}

Now we touch the issue about how the morphology of the crystalline domains depends on the temperature of the crystallizing film. To do this, we consider the critically-sized nuclei $N=n_{c}$ as well as the crystalline domains, whose size is $N_{cl}=200$ particles, and determine their morphological characteristics $R_{cl}$, $R_{v}$ and $d_{surf}$. For each the considered temperature states, values of these characteristics were obtained from averaging over results of ten independent simulations.
	
As discussed in Section 3.2, there is the crossover temperature $T\approx0.9\,\epsilon/k_{B}$ at which the linear size $r_{c}=[3N_{cl}/(4\pi\rho)]^{1/3}$, determined for $n_{c}$-sized nucleus, becomes comparable to half the width of the film. Remarkably, the same crossover temperature appears for the temperature dependencies of the characteristics $R_{cl}$ and $d_{surf}$, defined for the critically-sized nucleus [see Figure~\ref{fig_7}(a)]. For the thermodynamic states with the temperatures $T<0.9\,\epsilon/k_{B}$, the cluster size $R_{cl}$ increases slightly with temperature, whereas the thickness $d_{surf}$ does not change with temperature and takes the value equal to the thickness of one atomic layer. Then, for the temperatures above $T\approx0.9\,\epsilon/k_{B}$, we find that the cluster size increases with temperature mainly due to significant changes in the surface layer.

The crystalline domains of the size $N_{cl}=200$ particles have the linear sizes much larger than the width $L_{z}$ of the film, therefore, the evaluated quantities $R_{cl}$, $R_{v}$ and $d_{surf}$ have to characterize the morphology of the crystalline domains formed within the film plane. No effect due to the finite film thickness will appear for these structures. Thus, these large-sized domains have the width along $z$-axis equal to $L_{z}$ and their average linear size decreases with temperature due to structural changes along the $xy$-plane coincided with the film plane. Namely, we find that these structures become more ramified that is detected by the corresponded growth of values of the quantity $d_{surf}$ with temperature decrease [see Figure~\ref{fig_7}(b)] and that is in agreement with the results of Section 3.1.

\section{Conclusions}

The films with a characteristic thickness of less than $10$~nm are known as ultra-thin films and have been intensively studied since the 1980s. The ultra-thin films can be produced by means of the method of nanostructuring technology, the deposition methods and/or the electro-spinning technique~\cite{Zhong_Mao_2014,Wu_Pan_2007,Xue_Xia_2017,Liao_Fukuda_2014}. As known, even films whose size is $\sim1$~nm are experimentally achievable. In this regard it is appropriate to mention Refs.~\cite{Smith_1985,Murata_Sugiyama_1989}, where the experimental studies for gold films with thicknesses from $1$ to $100$~nm were reported. Further, taking into account that the size of gold atom is $\sigma\approx0.29$~nm, we find that the thickness of our simulated film for the case of gold will be $\approx1.39$~nm. Thus, the size of the simulated system is achievable for experimental methods, and the results of the present study can be experimentally verified. Moreover, Ekinci and Valles reported in Ref.~\cite{Ekinci_Valles_1998} experimental results related with a polycrystalline structure formation in ultra-thin Au and Pb films of the thickness $\leq1.6$~nm. The possibilities of modern experimental techniques make it possible to study even thinner samples for the case of alloys~\cite{Sun_Cao_2014}.

One of our main results is direct evidence that the crystallization of one-component system is initiated through the homogeneous crystal nucleation mechanism even at deep supercooling levels with the temperatures below than the glass transition temperature. The growing crystalline nuclei are detectable in the system at all the considered supercooling levels, whereas the waiting time and the critical size of the nuclei depend on the temperature. We find that the temperature dependence of the nucleation waiting time has a non-monotonic character similar to that is observed for the homogeneous crystal nucleation in bulk supercooled systems. Further, the critical size varies with the temperature according to the same law as in the case of the bulk crystallizing systems. This is supported by comparison with simulation results for Al, Ni, Cu$_{50}$Zr$_{50}$ and with experimental data for lithium disilicate. However, for the case of the film, the temperature dependence of the critical size has two regimes with the crossover temperature $T/T_g\approx1.15$ ($T\approx0.9\,\epsilon/k_{B}$), at which the linear size the nucleus becomes comparable with the width of the film.

The morphological features of the growing crystalline domains depend on the supercooling level of the crystallizing metallic film. Namely, we find that the liquid metallic film transforms into a monocrystal at low supercooling, whereas a polycrystalline structure is generated at deep supercooling levels. This follows directly from results of the cluster analysis and from inspection of the computed diffraction patterns.  All the crystalline domains are characterized by the linear scales associated with the size of the core spherical region, the size of the homogeneous part and the thickness of the surface layer of the domain. Evaluation of these characteristics displays that the critical size of the crystalline nucleus emergent in the metallic film takes different values at the various supercooling levels, and this is mainly due to the difference in the thickness of its surface layer. In addition, we find that the surface layer of the crystalline domains determines their specific morphology at the stage of their growth.

The results of the present work make a contribution to understanding of mechanisms of phase transformation in supercooled liquid and glassy systems as well as the results can be used to develop the methods of nanocrystallites fabrication with required structural and morphological properties. Further, the results of the given study can be extended as follows:

(i) As found, the structure of the crystallized metallic film is determined by supercooling level of the parent liquid sample. Therefore, it is reasonable to expect that these crystallized films will have the specific features related with the electron conductivity and the mechanical strength.

(ii) It would be interesting to investigate the surface properties of the crystallized films.

\section*{Acknowledgments}
The computational part of the work was supported by Kazan Federal University and RFBR according to the research projects No. 18-02-00407 and 18-32-00021. The main part of the work was supported by the subsidy of the Ministry of Education and Science of Russian Federation (grant No. 3.2166.2017/4.6) allocated to Kazan Federal University for performing the state assignment in the area of scientific activities.

\bibliographystyle{unsrt}

\end{document}